\documentclass[12pt]{iopart}
\usepackage{iopams}
\expandafter\let\csname equation*\endcsname\relax
\expandafter\let\csname endequation*\endcsname\relax
\usepackage{amsmath}
\usepackage{iopams}
\usepackage{graphicx}
\usepackage{epstopdf}
\usepackage{bm,epsfig}
\usepackage{amssymb}
\usepackage{latexsym}
\allowdisplaybreaks 
\begin{document}

 \title{Super-ultralow temperature laser cooling via interacting dark-state resonances}

 \author{Vase Moeini, Seyedeh Hamideh Kazemi and Mohammad Mahmoudi}

 \address{Department of Physics, University of Zanjan, University Blvd., 45371-38791, Zanjan, Iran}
\vspace{10pt}
\begin{indented}
\item[]{\today}
\end{indented}

\begin{abstract}
We propose a laser cooling mechanism that leads to a temperature significantly lower than the single-photon recoil limit, about $4\times 10^{-4}\,E_{r}$. This mechanism benefits from sharp and high-contrast spectra which are induced by interacting dark-state resonances. It is theoretically demonstrated that four-level atoms illuminated by two counter-propagating probe beams and two additional beams directed perpendicularly to other two, exhibit new cooling effects; For red detuned probe lasers, atoms can be subject to a strong viscous force with an extremely small diffusion, characteristic of heating caused by the stochastic nature of spontaneous emission processes. By quantum mechanical simulations, we then find that the lowest temperature approaches $0.3$ nK for the case of mercury, significantly lower than the recoil energy limit. A further advantage of our proposed scheme is that there is no need for an external magnetic field or a strong external confining potential.

\end{abstract}



%
%
%
%

 \section{Introduction}
Laser cooling and trapping have revolutionized the landscape of atomic physics and research in this area has been awarded a number of Nobel prizes during the last two decades \cite{chu,Tannoudji,Phillips,Ketterle,Cornell}. Some of the applications that have and will continue to benefit greatly from these advances are atom interferometry \cite{Kasevich}, atom lithography \cite{palm}, precision laser spectroscopy \cite{4}, quantum information processes \cite{1a}, measurement of ultracold chemical reactions \cite{10}, and study of few-body phase transitions \cite{14a,15a,17a,18a}. 

It is now more than four decades since H\"{a}nsch and Schawlow realised that narrow linewidth lasers could exert a significant force on atoms and could be used for cooling \cite{Schawlow,note}. This proposal, the first cooling mechanism for neutral atoms, started with a low-density gas interacting with a single laser beam. With lasers confined to the lower half of the Doppler linewidth, only those atoms which are moving towards the laser source find the Doppler-shifted up in frequency so that they can lose energy and momentum. Investigating competition between cooling and heating due to the stochastic nature of the spontaneous emission process, one can find the equilibrium temperature which is limited by the natural width of the excited states. For a typical atom, this temperature, known as Doppler temperature, is of order of $10^{-4}$K \cite{adams}. It soon appeared that such a limit can be beaten \cite{lett} and after that, laser cooling to temperatures below the Doppler limit has been an area of much activity, both in experiment and in theory. One key feature of the models, distinguishing from the earlier works on the Doppler cooling, relates to the fact that the multi-level structure of atoms plays a substantial role in cooling processes \cite{Dalibard,Ungar}. Many experiments have been conducted on diverse configurations leading to sub-Doppler cooling, among which are polarization gradient cooling \cite{lett,weiss2}, velocity selective magnetic resonance cooling \cite{Hoogerland}, and cooling by Raman resonances \cite{Kasevich2}. For more comprehensive listing and description of the configurations, the reader is referred to \cite{Bergeman,Metcalf}.  

On the other hand, the phenomenon of dark-state or coherent population transfer \cite{Arimondo}, a well-known concept in quantum optics and laser spectroscopy, forms the base for a wealth of important effects, such as electromagnetically induced transparency (EIT) \cite{eit}, lasing without inversion \cite{scully}, femto-second generation \cite{hariss2}, and slow light \cite{harris,kash,budker}. In this context, sub-recoil laser cooling methods have been proposed which take advantage of the dark-state resonances \cite{Aspect,15c,16c,he}. Typically the recoil limit corresponds to a temperature of between $10^{-7}$ and $10^{-6}$K. A particular instance of these methods is a scheme based on velocity-selective optical pumping of atoms into a nonabsorbing coherent superposition of states, leading to transverse cooling of $^4 $He atoms to a temperature lower than both usual Doppler cooling limit and the one-photon recoil energy \cite{Aspect}. It is imperative to point out a remarkable work that suggested the use of spectral feature generated by an EIT in a Lambda-type three-level system for approaching a temperature around the single-photon recoil energy \cite{he}. In spite of the pronounced success, there still exists a continuing need for lower temperatures than conventional laser cooling techniques can provide. 

Double-dark resonance (DDR), a novel spectral feature appearing in a system with multiple coherent interacted superposition states, was first studied in 1999 \cite{lukin}. The coherent interaction leads to the splitting of the dark states and the emergence of very sharp, high-contrast structures in the optical spectra. Two years later, quantum interference induced by the interacting dark-state resonances was experimentally observed \cite{chen2001} and after that several different schemes were explored for Doppler-free resonance \cite{ye}, nonlinear optics \cite{yelin}, and group velocity controlling \cite{mahmoudi}. In this paper, we present a laser cooling technique based on DDR in a four-level atomic system, which has allowed us to achieve cooling of atoms to temperatures as low as $4\times 10^{-4}\,E_{r}$, extremely far below the recoil energy limit ($E_{r}$). Physically, the attainable temperature in laser cooling methods is determined by a balance between dissipative force and  heating rate due to the spontaneous emission. Here, distinctive spectral feature of the DDR significantly reduces the heating rate and simultaneously strengthens the dissipative force and finally leads to a surprisingly low temperature of about $0.3$ nK for the case of mercury, which may hold great promise for practical applications in laser cooling. We must reiterate the importance of the fact that the ultralow-temperature can be achieved without a magnetic field or a strong confining potential.

\section{The Model}
\begin{figure}
\centering
\includegraphics[width=0.8\linewidth]{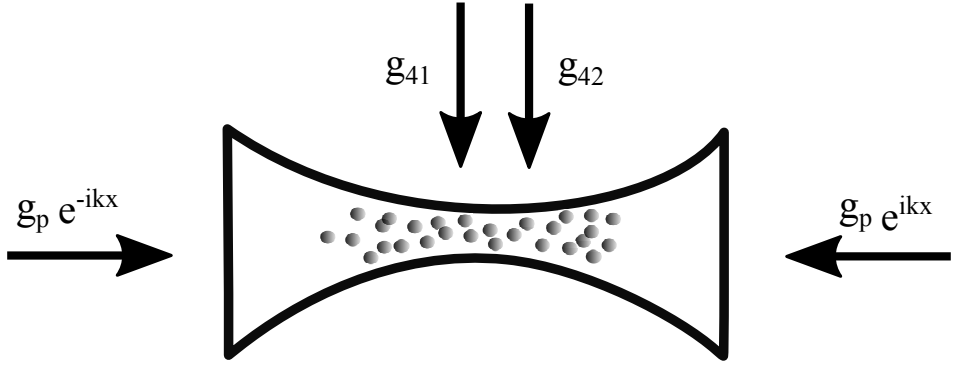}
\caption{ Two counter-propagating probe laser beams with equal intensity interact with four-level atoms which are considered to be transversely confined so that they restricted to move along x direction. The coherent and the coupling fields- $g_{42}$ and $g_{41}$, respectively- propagate perpendicularly to the probe beams.}
\label{figure1}
\end{figure}

\begin{figure*}
\centering
\includegraphics[width=0.9\linewidth]{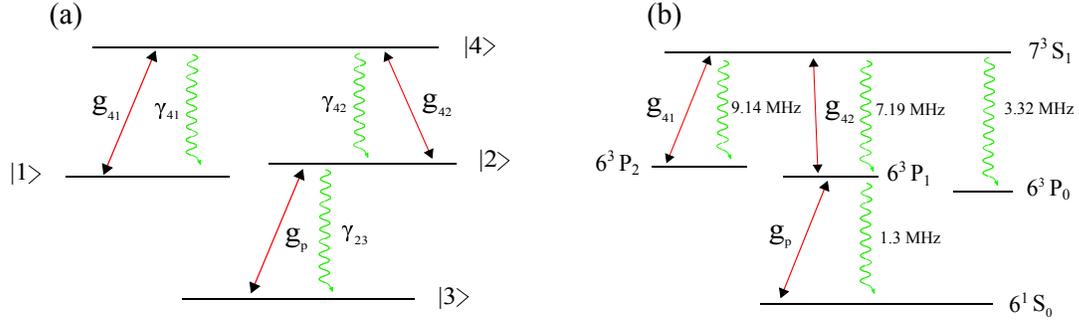}
\caption{(a) Considered energy scheme of the four-level atomic system in which wavy lines show the spontaneous decays from the excited states. (b) illustrates a possible realization of the scheme in mercury. Noting that the population transferred to state $6^3 P_0$ has to be compensated via a repump field.}
\label{figure2}
\end{figure*}

The considered atomic system is composed by four-level atoms which are coupled by laser fields and restricted to move along x direction, according to the scheme depicted in Fig.~\ref{figure1} and Fig.~\ref{figure2}(a). A strong coherent field with frequency $\omega_{42}$ is applied to transition $ \vert 2\rangle - \vert 4\rangle$ with Rabi frequency $g_{42}$. A weak coupling field with Rabi frequency $g_{41}$ and frequency $\omega_{41}$ couples to the transition $ \vert 1\rangle - \vert 4\rangle$. Moreover, $ \vert 3\rangle - \vert 2\rangle$ transition is driven by a weak probe field with frequency $\omega_{23}=\omega_{p}$ and Rabi frequency $g_p$. The spontaneous decay rates on the dipole-allowed transitions are denoted by $ \gamma_{41}$, $ \gamma_{42}$, and $ \gamma_{23}$. We also define $ E_k\, [k \in \lbrace1, ..., 4\rbrace]$ as the energies of the involved states and the transition frequencies are denoted by $\overline{\omega}_{ij}= (E_{i}-E_{j})/ \hbar \, [i \in {2,4}, j \in \lbrace1, ..., 3\rbrace] $. Moreover, laser field detuning with respect to the atomic transition frequency is given by $\Delta_{ij}= \omega_{ij}-\overline{\omega}_{ij}$. Noting that the general expression for a Rabi frequency is defined as $g=(\vec{\mu} . \vec{E})/{\hbar}$ with $\vec{\mu}$ and $\vec{E}$ being the atomic dipole moment of the corresponding transition and amplitude of the field, respectively. This system can be realized in mercury with the probe transition at 253.7 nm (see Fig.~\ref{figure2}(b) for more details). 

As the prominent work by Holland \textit{et al.} \cite{he} argues, in a case of a weak probe field, the cooling mechanism is similar to a Doppler cooling; A moving two-level atom is illuminated by two beams counter-propagating in the $\pm$ x in which the probability for the atom absorbing light depends on its velocity through the Doppler effect. For the atom moving to the right, a Doppler-shifted frequency from the left beam, i.e., $\omega_{p} + \mathrm{k} \upsilon$ is obtained and from the right propagation beam, a frequency of $\omega_{p} - \mathrm{k} \upsilon$. If the light is red-detuned, the atom is most likely to absorb photons moving towards. Similar to the conventional Doppler cooling of the two-level atom, we here adopt a similar approach and assumed that the probe field is composed of a pair of counter-propagating laser beams with equal frequency aligned in x direction and z-polarized two other beams propagate along the y direction (see Fig.~\ref{figure1}). So, the electric field in the z direction can be written as 
\begin{equation}
E_{\mathrm{z}}= 2 E_{p} \cos(\mathrm{kx}) e^{i \omega_{p} t} + E_{41} e^{i \omega_{41} t} + E_{42} e^{i \omega_{42} t} + \mathrm{c.c},
\end{equation}
with $\mathrm{k}= \omega_{p}/c$ being the wave vector of the probe field. 

Hamiltonian of the system for the special case $\Delta_{41}=\Delta_{42}=0$ and $\Delta_{23}=\Delta_{p}$ as the probe field detuning can be written as 
\begin{equation}
\hat{H}= \frac{\hat{\mathrm{p}}^2}{2 m} + \hbar \Delta_{p} \hat{\sigma}_{33} + \hbar [2 g_{p} \cos(\mathrm{k\hat{x}}) \hat{\sigma}_{32} + g_{41} \hat{\sigma}_{14} + g_{42} \hat{\sigma}_{24} + \mathrm{H.c.} ].
\end{equation}
Here H.c. corresponds to the Hermitian conjugate of the terms explicitly written in the Hamiltonian. The first term accounts for atomic motion and other terms describe the internal atomic levels. We also introduce the quantum projection operator $\hat{\sigma}_{lm}= \vert l \rangle \langle m \vert$.

In the presence of dissipation, the evolution of the system is described by Born-Markov quantum master equation for the density matrix $\hat{\rho}$
\begin{eqnarray}
\frac{d \hat{\rho}}{dt} &=& \frac{i}{\hbar}[\hat{\rho},\hat{H}] + \gamma_{23} \int du N_{1}(u)    
\mathcal{L}[\hat{\sigma}_{32} e^{i\mathrm{k}u\hat{\mathrm{x}}}] \hat{\rho} + \gamma_{42} \int du N_{2}(u)    
\mathcal{L}[\hat{\sigma}_{24} e^{i\mathrm{k}^{\prime}u \hat{\mathrm{x}}}] \hat{\rho}  \nonumber \\
&+& 
\gamma_{41} \int du N_{3}(u)    
\mathcal{L}[\hat{\sigma}_{14} e^{i\mathrm{k}^{\prime\prime}u \hat{\mathrm{x}}}] \hat{\rho}.
\label{eq0}
\end{eqnarray}
Where the Linblad superoperator $\mathcal{L}[\hat{O}] \, \hat{\rho} = 1/2 ( 2 \hat{O} \, \hat{\rho} \, \hat{O}^{\dagger} - \hat{O}^{\dagger} \hat{O} \hat{\rho} - \hat{\rho} \, \hat{O}^{\dagger} \, \hat{O} )$ describes incoherent processes and $\mathrm{k}^{\prime}$, $\mathrm{k}^{\prime\prime}$, and $N_{1,2,3}(u)$ are the wave-vector associate with $ \vert 2\rangle - \vert 4\rangle$, $ \vert 1\rangle - \vert 4\rangle$, and the corresponding normalized dipole radiation pattern projected along the x direction, respectively.

For temperature sufficiently high as to satisfy that the typical kinetic energy ($E_r$) exceeds the recoil one, the motional dynamics is well described by a semiclassical treatment in which the translational motion is treated classically, the internal motion quantum mechanically and mapping $\langle \hat{\mathrm{x}} \rangle \rightarrow \mathrm{x}$ and $\langle \hat{\mathrm{p}} \rangle \rightarrow \mathrm{p}$ is made with $\mathrm{x}$ and $\mathrm{p}$ being as classical variables. We then proceed to calculate two prominent parameters to characterize a sample of atoms in the semiclassical approximation, i.e., friction coefficient and diffusion constant.

The force on the atom is given by the negative gradient of the Hamiltonian which takes the form
\begin{eqnarray}
F=\frac{d\mathrm{p}}{dt}= \langle \frac{d \hat{\mathrm{p}}}{dt} \rangle =- \langle \frac{\partial \hat{H}}{\partial \hat{\mathrm{x}}}\rangle  = 4 \hbar \mathrm{k} g_{p} \sin(\mathrm{kx})  \mathrm{Re}[\rho_{32}].
\label{eq3}
\end{eqnarray}
As is seen, the main observable is the coherence, $\rho_{32}$, which can be obtained by solving the density matrix equations of motion which have been worked out in Ref. \cite{mahmoudi}. 

On the other hand, the linear susceptibility of the weak probe field can be written as \cite{scully}
\begin{equation}
\chi (\Delta_{p})= \frac{N \eta_{p}}{\epsilon_0 E_p} \rho_{23}(\Delta_{p}),
\end{equation}
where $N$ and $\eta_{p}$ are, respectively, atom number density in the medium and the probe transition dipole moment. Real and imaginary parts of the susceptibility correspond to the dispersion and absorption, respectively. Notice that, throughout the discussion, for simplicity we have written relation between the coherence and the susceptibility as $\chi (\Delta_{p})= \rho_{23} (\Delta_{p})/ g_p$. 

Next, we derive expression for $\rho_{23}$ when the atom is not at rest, but moving with velocity $\upsilon = \mathrm{p}/m$. In the weak probe regime, modifying the equation for the coherence to incorporate atomic motion, we find
\begin{equation}
\rho_{23} (\Delta_{p}) \approx g_p [ e^{-i \mathrm{k x}} \chi (\Delta_p - \mathrm{k} \upsilon) + e^{i \mathrm{k x}} \chi (\Delta_p + \mathrm{k} \upsilon)],
\label{eg10}
\end{equation}
and by substituting Eq.~(\ref{eg10}) into Eq.~(\ref{eq3}), in the limit of slowly moving atoms, we find
\begin{eqnarray}
F &\approx & 4 \hbar \mathrm{k} \vert g_p \vert ^2 \sin(2 \mathrm{k x}) \mathrm{Re}[\chi (\Delta_p)] \\ \nonumber
&-&
4 \hbar \mathrm{k}^2 \upsilon \vert g_p \vert ^2 [1- \cos(2 \mathrm{k x})] \partial_{\Delta_p} \mathrm{Im}[\chi (\Delta_p)].
\end{eqnarray}
Separating out the real and imaginary parts of $\chi$, it is found that there are contributions from the reactive and dissipative components of the atomic response, so terms appeared can be interpreted as the conservative dipole force and radiation one, respectively. The former can be ignored in the weak probe field and averages to zero over a wavelength, while the latter will give rise to a velocity-dependent dissipative force which is approximately linear with respect to the atom velocity in the sense that negative of the slope yields the friction coefficient $\eta(\mathrm{x})$ \cite{he}. Note that the friction coefficient can be easily earned and averaged over one wavelength via the steady-state solutions of the coherence $\rho_{23}$. 

In order to accurately incorporate the role of the fluctuations, we then include a stochastic term in the equation of motion 
\begin{equation}
\frac{d \mathrm{p}}{dt}= - \eta \mathrm{p} + \hat{\xi}(t),
\label{langevin}
\end{equation}
where $\hat{\xi}(t)$ is the classical noise satisfying $\langle \hat{\xi}(t) \rangle  =0 $ and its two-time expection value is defined by $ \langle \hat{\xi}(t) \hat{\xi}(t^{'}) \rangle = 2 D \, \delta(t-t^{'}) $ with $D$ being the diffusion constant as follows $ 2 D= \langle d \hat{\mathrm{p}}^2/dt  \rangle - 2 \langle \hat{\mathrm{p}} \rangle \langle d \hat{\mathrm{p}}/dt  \rangle$ \cite{zoller}. 

Generally, there are two main heating sources; We can understand the first one in terms of the momentum kicks imparted to atoms by spontaneous emission and the second arises from the zero-point fluctuations of the atomic dipole moment \cite{mckay,he}. Indeed, we include the stochastic term ($\hat{\xi}(t)$) to account for both heating sources. Physically, the finite equilibrium temperature $T$ is determined by the balance of dissipative force and diffusion and consequently, we expect that cooling will proceed until the heating rate equals the cooling one.

We have then taken the dipole pattern: $N_{1,2,3}(u) = 1/2 \delta (u+1) + 1/2 \delta (u-1)$ which is explained by assuming that the momentum kicks happen with equal probability and only along the $\pm \mathrm{x}$. Now, we can calculate the diffusion constant by solving the master equation for the density matrix, Eq.~(3), and following the prescription in Ref. \cite{simon}. Assuming that the system approaches thermal equilibrium at long time ($\eta t \gg 1$), the final temperature attainable by the cooling mechanism is determined by the balance of the diffusion constant and friction coefficient $k_{B} T = D /m \eta$.

Using the semiclassical approach described in previous paragraphs, we will calculate friction coefficient, diffusion constant, and thus the equilibrium temperature for the four-level atomic system of mercury. Our results are represented in scaled quantities to obtain the best possible comparison with other cooling techniques; Friction coefficient and diffusion constant are divided by $E_r/\hbar$ and $m E_r$, respectively. The temperature is also scaled by $E_r/k_B$. Noting that the recoil velocity $\upsilon_r$ is the change in the atom velocity when absorbing or emitting a resonant photon, and is given by $\upsilon_{r}=\hbar k_{L}/m$ and accordingly, the recoil energy is defined as the kinetic energy of an atom moving with velocity $\upsilon=\upsilon_{r}$, which is $\hbar \omega_{r}=\hbar^2 k_{L}^2/2m$.

\section{Results}

We will break the analysis into two cases: 1) Without interacting dark-state resonance in which we have an Autler-Townes doublet with a dip in the absorption at zero detuning, i.e., a partial EIT. 2) With interacting dark-state resonance where a high-resolution peak appears in the optical spectra due to the presence of interacting dark resonances. In order to orient our discussion, we begin with the case of the Autler-Townes doublet and by calculating the friction coefficient and diffusion constant as well as final temperature from semiclassical treatment, show that the scheme exhibits behavior similar to that found in previous studies in which the lowest achievable temperatures have been limited to around the single-photon recoil energy. We then discuss the case of the DDR, a three-photon phenomenon, and show how the lowest temperature can be orders of magnitude smaller than the recoil energy. Noting that, here, two different approaches are used in order to determine the achievable temperature: In addition to a semiclassical approach that allows us to derive the friction coefficient, the diffusion constant and the equilibrium temperature, we use a full-quantum one.

\subsection{Without interacting dark-state resonance}
\begin{figure}[t]
\centering
\includegraphics[width=0.7\linewidth]{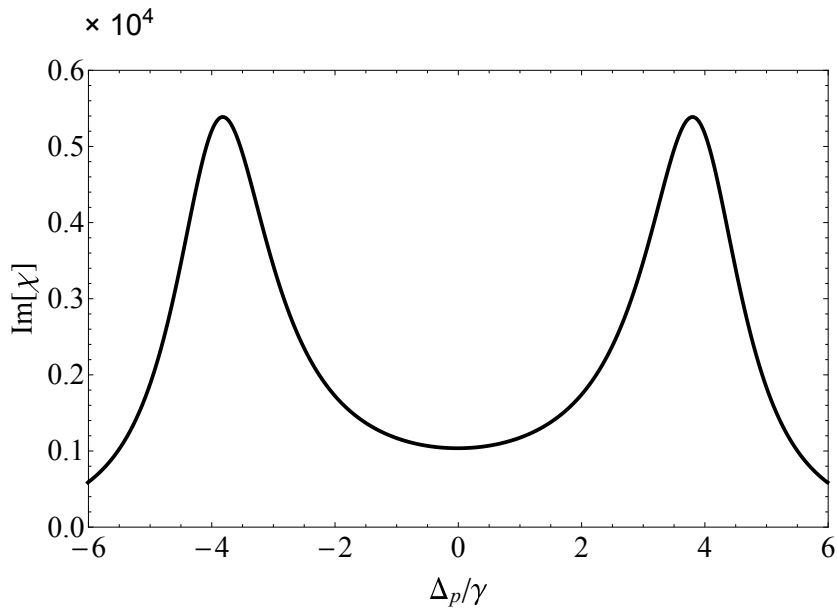}
\caption{ Imaginary part of the susceptibility $\chi$ as a function of the probe detuning $\Delta_{p}$ for the parameters  $\gamma_{41}=\gamma$, $\gamma_{23}=0.14 \gamma$, $\gamma_{42}=0.79 \gamma$,  $\gamma_{13}=0.01 \gamma$, $g_{p}=10^{-4} \gamma$, $g_{41}=0$, $g_{42}=4 \gamma$, and $\Delta_{42}=\Delta_{41}=0$. }
\label{figure3.1}
\end{figure}

In Fig.~\ref{figure3.1}, we show the imaginary part of the probe field susceptibility $\chi$ versus the probe detuning $\Delta_{p}$, which corresponds absorptive properties of the medium. In this figure, the perturbing laser field is switched off ($g_{41}=0$) and other parameters are $\gamma_{41}=\gamma$, $\gamma_{23}=0.14 \gamma$, $\gamma_{42}=0.79 \gamma$, $\gamma_{13}=0.01 \gamma$, $g_{p}=10^{-4} \gamma$, $g_{42}=4 \gamma$, and $\Delta_{42}=\Delta_{41}=0$. It is worth noting that the ratios of the decay rates correspond to the case found in mercury. We have added a weak decay rate $\gamma_{13}$, since otherwise in the steady-state all population is trapped in $\vert 1 \rangle$. The driving field, with the Rabi frequency $g_{42}$, leads to an Autler-Townes doublet
with a dip in the absorption at zero detuning, i.e., partial EIT. In the case of a long-lived state $\vert 4 \rangle$, the EIT leading to the partial transparency would be more pronounced such that the absorption vanishes at zero detuning. 

\begin{figure}
\centering
\includegraphics[width=0.8\linewidth]{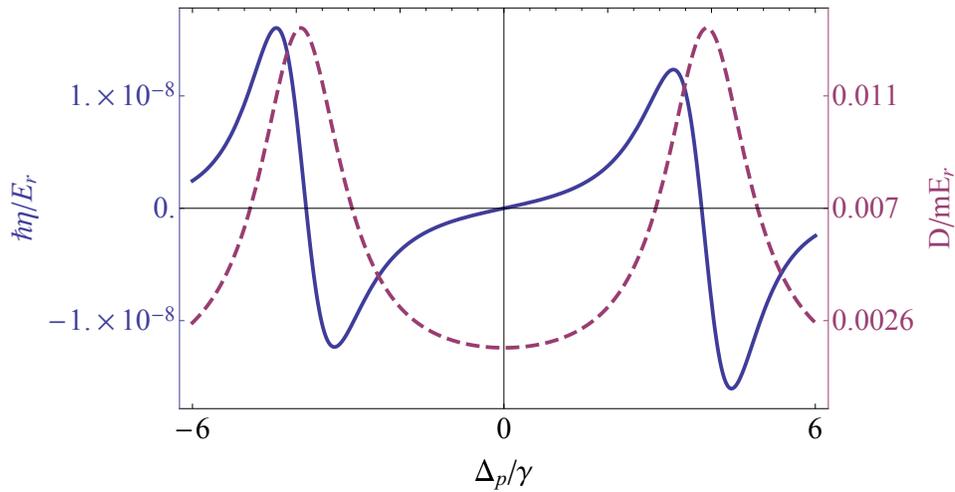}
\caption{ Friction coefficient (solid line) and diffusion constant (dashed line) as a function of the probe detuning $\Delta_{p}$. The parameters are the same as for Fig.~\ref{figure3.1}. }
\label{figure4.1}
\end{figure}

Fig.~\ref{figure4.1} shows friction coefficient and diffusion constant as a function of the probe detuning ($\Delta_{p}$). As is seen, we find a positive friction coefficient for the left side of the left peak and the right side of the right peak, meaning that the force will act to cool the atoms. In addition to the friction, however, we also need to calculate the diffusion, characteristic of heating caused by the stochastic nature of spontaneous emission processes. Hence, for considering the final temperature attainable by Doppler cooling, we should look at the diffusion (dashed line in Fig.~\ref{figure4.1}). 

\begin{figure}[!hb]
\centering
\includegraphics[width=0.7\linewidth]{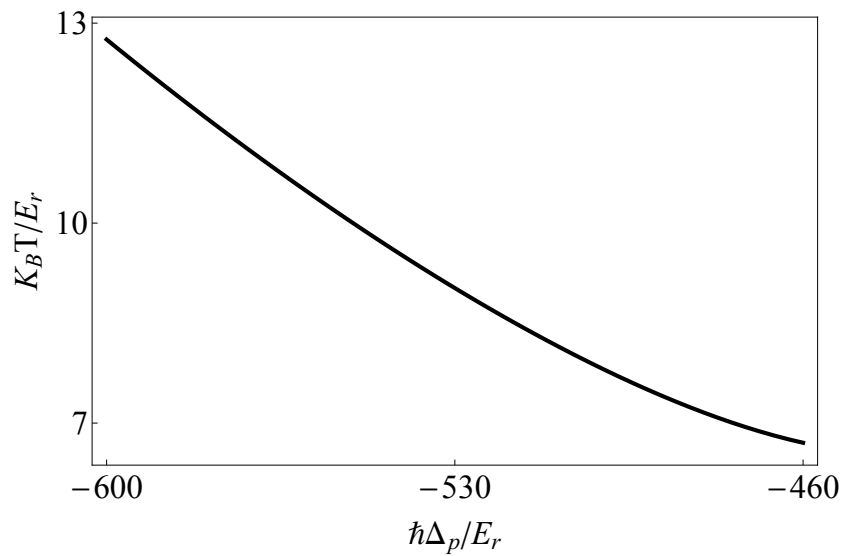}
\caption{ The final temperature calculated from the semiclassical treatment as a function of the scaled probe detuning ($\hbar \Delta_{p}/E_r$). The parameters are the same as for Fig.~\ref{figure3.1}.}
\label{figure5.1}
\end{figure}   

Since we now know both the diffusion constant and the friction coefficient, we can calculate the final temperature attainable by the Doppler cooling, which is shown in Fig.~\ref{figure5.1}. As is seen, the attainable temperatures are limited by the recoil limit and the lowest temperature is found to be 6.6 $E_r$ at $\Delta_{p}=-460\, E_r/\hbar$. The temperature is approximately the same value previously found in the sub-Doppler laser cooling where the lowest achievable temperatures have been limited to around the single-photon recoil energy. As an explicit example, consider the mercury and the final temperature would be 0.25 $\mu$K at $\Delta_{p}=-$45 MHz. Keeping in mind that the motion is well described by a semiclassical treatment as the temperatures exceeds the recoil limit and there is no need to adopt a fully quantum mechanical treatment of the motional wave-function.

\subsection{With interacting dark-state resonance}

\begin{figure}[b]
\centering
\includegraphics[width=0.7\linewidth]{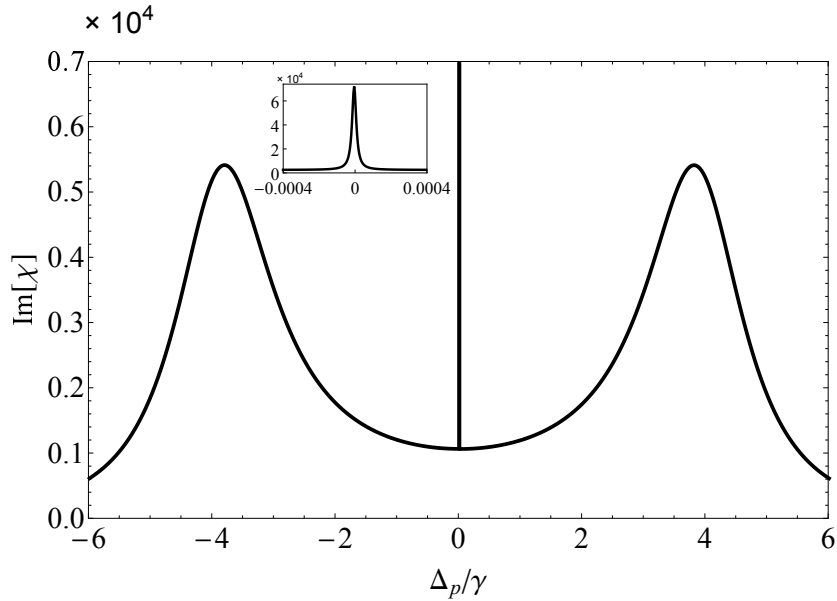}
\caption{ Imaginary part of the susceptibility $\chi$ as a function of the probe detuning $\Delta_{p}$ for $g_{41}=0.04 \gamma$ and $\gamma_{13}=0$. The other parameters are the same as in Fig.~\ref{figure3.1}. Inset shows a close-up on the central part of the curve.  }
\label{figure3}
\end{figure}

Here, we put very few restrictions on the scheme presented in previous subsection: A weak perturbing field with the Rabi frequency $g_{41}=0.04 \gamma$ is applied and a negligible decay on transition $\vert 3 \rangle \leftrightarrow \vert 1 \rangle $ is assumed, since a trapping in this state is now avoided by the additional laser field. Imaginary part of the probe field susceptibility versus the probe detuning $\Delta_{p}$ is plotted in Fig.~\ref{figure3}. The results are identical to Fig.~\ref{figure3.1} except for a narrow absorption spike at around zero detuning. Again, for a long-lived state $\vert 4 \rangle$, the transparency regions on each side of the absorption spike would become two points of EIT, i.e., a double dark state \cite{lukin}. In addition to the Autler-Townes doublet structure, we have a narrow absorption spike appeared around zero detuning due to the three-photon resonance $\vert 1 \rangle \rightarrow \vert 4 \rangle \rightarrow \vert 2 \rangle \rightarrow \vert 3 \rangle $. Noting that the width of the spike is much less than the natural linewidth.

\begin{figure}[t]
\centering
\includegraphics[width=0.8\linewidth]{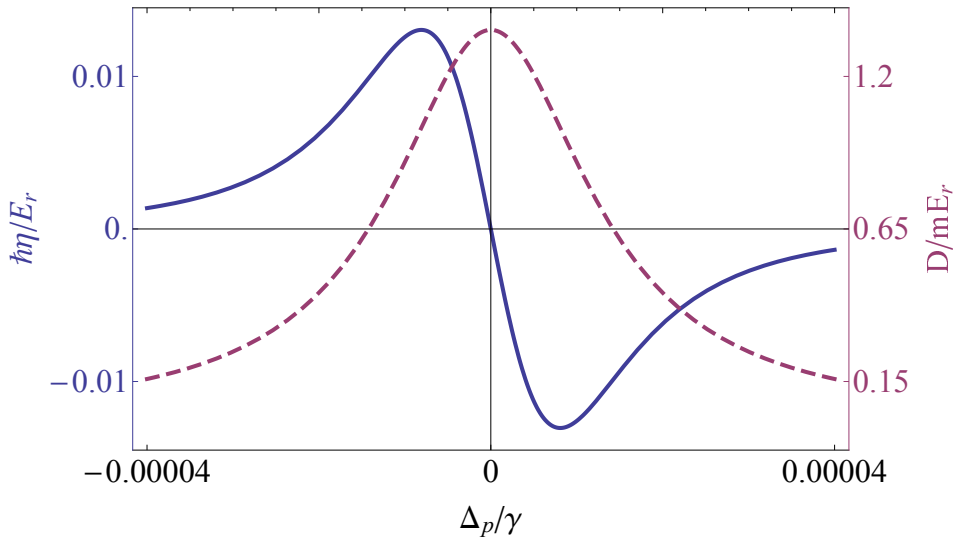}
\caption{ Friction coefficient (solid line) and diffusion constant (dashed line) as a function of the probe detuning $\Delta_{p}$. The parameters are the same as for Fig.~\ref{figure3}. }
\label{figure4}
\end{figure}

\begin{figure}[t]
\centering
\includegraphics[width=0.7\linewidth]{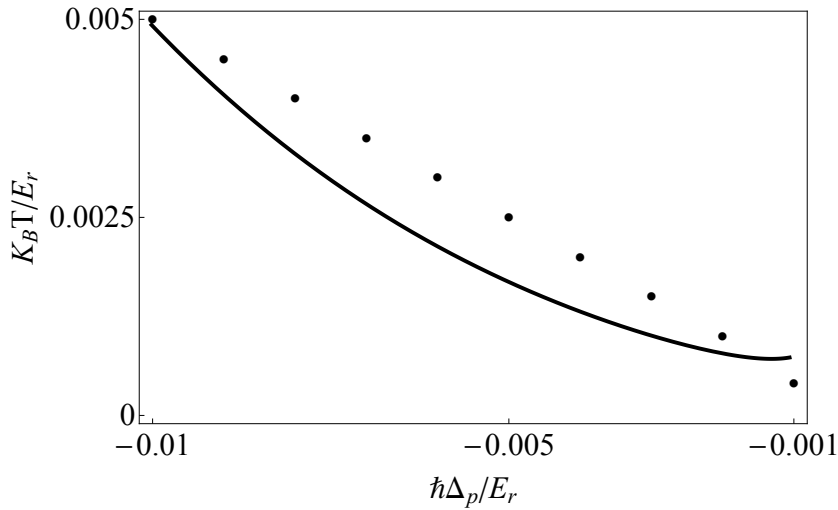}
\caption{ Calculation of final temperature as a function of the scaled probe detuning ($\hbar \Delta_{p}/E_r$). Semiclassical results (solid line) and MCWF numerical results (circle) with truncated momentum basis ranging from $-\mathrm{N}\hbar \mathrm{k}$ to $\mathrm{N} \hbar \mathrm{k}$ are presented. The parameters are the same as for Fig.~\ref{figure3}. The figure shows anticipated discrepancies between the solutions based on the semiclassical approach and the results from MCWF as $k_{B} T\ll E_{r}$. }
\label{figure5}
\end{figure}

The cooling in this case exhibits distinctly different characters, having a small heating rate and simultaneously a strong dissipative force, as will be shown below. Friction coefficient and diffusion constant as a function of the probe detuning are plotted in Fig.~\ref{figure4}. Similar to the conventional Doppler cooling of the two-level atoms, we here have a absorption peak but with a significant narrower linewidth, and by analogy, it might reasonably be expected to be necessary for the probe beam to be detuned to the red (lower frequency) of the resonance. So, for red detuning, $\Delta_p < 0$, the friction coefficient is positive which means that the system will cool. The important point to note is that the DDR may hold the potential to drastically strengthen the friction coefficient. Also, compared to a recent work on sub-Doppler laser cooling \cite{he}, heating rate and subsequently the diffusion constant are significantly reduced by about four order of magnitude mainly due to small upper-level populations. Using the diffusion and friction results, it is straightforward to find reachable temperature from the semiclassical approach which is plotted as solid line in Fig.~\ref{figure5}. 

As mentioned before, semiclassical treatment is valid only if the predicted thermal energy from the semiclassical treatment ($k_{B} T$) exceeds the recoil energy ($E_{r}$), in other words, if the atomic wave packet of the center-of-motion is well located within the region of the laser wavelength; otherwise, the semiclassical analysis is no longer valid and the degrees of freedom (position and momentum) should be treated as operators \cite{cohen}. As can be seen from solid line in Fig.~\ref{figure5}, for the whole detuning range of the pump laser, the predicted thermal energy from the semiclassical treatment is much smaller than the recoil energy. Accordingly, the coherence length of the atomic wave packets, $\xi \sim \hbar/m\Delta\nu$ with $\Delta\nu \sim \sqrt{k_{B} T /m}$, is much smaller than typical laser wavelength for laser cooling and the temperature should be found by solving the density matrix quantum mechanically.

In the following, we present the results of full quantum analysis using Monte-Carlo wave function (MCWF) method \cite{molmer} in which fluctuations and dissipation originate from a quantum jump. This approach, equivalent to the standard master equation one, can be applied to a wide variety of quantum optics problems \cite{he,castin}. For a system with a number of states much longer than unity, the MCWF approach can reduce computation complexity compared with the master equation treatment as the wave function involves only N components, smaller than the number of variable involving in the density matrices. Consider a wave-function that is expanded in terms of the internal states (i) and quantized momentum ($\mathrm{p}^{\prime}$). At a time $t$, the general quantum state in this basis can be written as
\begin{equation}
\vert \psi(t) \rangle = \sum\limits_{\mathrm{p}^{\prime}_{n}=-\mathrm{N} \hbar \mathrm{k}}^{\mathrm{N} \hbar \mathrm{k}}    \sum\limits_{i=1}^{3}  c_{i,n}(t) \vert i,\mathrm{p}^{~}+\mathrm{p}^{\prime}_{n} \rangle.
\end{equation}
Indeed, we expand the center of mass components of the wave-function on a set of states with momenta $\mathrm{p}^{~}+n \hbar \mathrm{k}$ with integer $n$, giving a gird of momentum basis states that ranges from $-\mathrm{N}\hbar \mathrm{k}$ to $\mathrm{N} \hbar \mathrm{k}$. In practice, we have to limit the  grid, and we consider it sufficient to take $-50 \leq n \leq 50$ in such a way that a further increase of $\mathrm{N}$ did not give the different results. Generally, the evolution of $c_{i,n}(t)$ consists of sequences of two steps; First, the wavefunction evolves with the non-Hermitian Hamiltonian  $H_{eff}= H -i \hbar \gamma_{23} \hat{\sigma}_{22} - i \hbar (\gamma_{41}+\gamma_{42}) \hat{\sigma}_{44} $ in order to account for the dissipative processes induced by the spontaneous emission. Second, we randomly decide whether a quantum jump occurs. In fact, when the norm of the wavefunction is smaller than a fixed initially produced random number, this single dissipative event occurs.

Finally, we use the MCWF method to accurately determine the minimum reachable temperature. Here, a comparison is made of the solutions based on the semiclassical approach (solid line) and the MCWF treatment of the full quantum solutions (circle). As depicted in Fig.~\ref{figure5}, the temperature does go far below $ E_{r}$ and obtained sub-Doppler temperature ($\sim 1.5\, E_r$) in the recent work of Holland \textit{et al.} is surpassed \cite{he}. Remarkably, the lowest temperature is found to be about 4 $ \times 10^{-4} \,E_{r}$ at $\Delta_p=-0.001\,E_{r}/\hbar$. Indeed, our suggested laser cooling mechanism takes advantage of the DDR, a phenomenon due to three-photon resonances, and cooling of atoms down to such surprisingly low temperatures can be achieved. While, in the above-mentioned mechanism based on the EIT, e.g., a two-photon phenomenon, the temperature does not go below $1.5\, E_{r}$: A reason for substantiate the claim that, under the appropriate circumstances, the transition including more number of photons, the lower minimum reachable temperature.

In the suggested configuration with mercury and by considering the typical values, the final cooling temperature at $\Delta_{p}$= -97 Hz is estimated to be 0.3 nK. It is also of interest to see whether the spontaneous decay rates have a substantial effect on the cooling, so we have also calculated the minimum reachable temperature for the commonly used alkali atom $^{87}$Rb; Considering its typical values \cite{he}, the final cooling temperature is estimated as 0.95 nK. We attribute this behavior to the fact that the DDR now becomes slightly broader for the $^{87}$Rb and  thus the temperature does not reach as low as that for mercury.

 



Our discussion thus far implicitly assume that the atomic gas was dilute, while in laser cooling experiments, this approximation will fail and the sample is dense enough that the atoms start to affect each other. So, we would need to account for two effects: Laser attenuation and multi-atom interactions. 1) A thick sample attenuates any laser beam propagating through it and subsequently the intensity of the cooling beams will shrink with distance and may be subject to strong nonlinear optical effects \cite{childress}. Therefore, the force will acquire a spatial gradient resulting in position-dependent cooling and so in real systems, fewer atoms may be subject to the dissipative force. This adverse effect, arising from the photon absorption and dispersion in the sample, may considerably affect the achievable temperature. Although, using the almost zero-absorption region on the DDR structure could significantly lessen this unfavorable effect. 2) A more prominent effect emerges when we consider the multi-atom interactions in the dense gas. Indeed, these interactions including ground-state collisions of atoms, induced electric dipole-dipole interactions as well as light-assisted collisions could limit the total number of atoms that can be cooled \cite{he}. In many cases, these kind of interactions can be mediated by photon emission and so can be suppressed by using the almost negligible absorption region on the DDR structure \cite{he,chang}. We do not calculate the multi-atom effects in current paper, however, the multi-atom Heisenberg-Langevin equation or, alternately, a multi-atom master equation \cite{gross} as well as coupled atom dynamics in a dense sample \cite{castin2} would provide a valuable extension for future work.  

As a last remark, we would like to point out that we deal with temperatures near or below Bose-Einstein condensate (BEC) critical one; Typical critical temperature of the BEC is about a few hundred nK, almost two orders of magnitudes higher than the cooling limit achieved in our suggested scheme. For instance, the final cooling temperature is found to be about 1 nK for the rubidium atoms. For those atoms at 1 nK, the de Broglie wavelength of the atom is about 6 $\mu$m which is almost comparable to the atomic cloud size, resulting in modifying the efficacy of the cooling or altering the attainable temperatures. Nonetheless, there is one outstanding feature about the suggested system, and indeed any system with the DDR, and that is significantly reducing the heating rate and simultaneously strengthening the dissipative force.




\section{Conclusion}

In summary, this paper has proposed and analyzed a potentially super-ultralow temperature laser cooling mechanism in a four-level atomic system which takes advantage of the sharp spectra induced by the DDR. We developed the results of a semiclassical approach and found that the distinctive spectral feature of the DDR significantly reduces the heating rate and simultaneously strengthens the dissipative force, compared to related works on sub-Doppler laser cooling. By treating both internal and translational degrees of freedom quantum mechanically, we then found that the scheme has allowed us to achieve cooling of mercury atoms to a surprisingly low temperature of about $0.3$ nK which could be orders of magnitude less than the best previously published results on sub-recoil laser cooling.

\section*{Funding}
S.H. Kazemi acknowledges the financial support of the Iran National Science Foundation (96008805).  

\section*{Acknowledgments} 
V.M. Moeini is grateful for valuable discussions with Peiru He.


\section*{References}

 \begin{thebibliography}{}
\bibitem{chu}
S. Chu, Rev. Mod. Phys. \textbf{70}, 685 (1998).  
\bibitem{Tannoudji}
C. N. Cohen-Tannoudji, Rev. Mod. Phys. \textbf{70}, 707 (1998).
\bibitem{Phillips}
W. D. Phillips, Rev. Mod. Phys. \textbf{70}, 721 (1998).
\bibitem{Ketterle}
 W. Ketterle, Rev. Mod. Phys. \textbf{74}, 1131 (2002).
\bibitem{Cornell}
E. A. Cornell and C. E. Wieman, Rev. Mod. Phys. \textbf{74}, 875 (2002).
\bibitem{Kasevich}
M. Kasevich and S. Chu, Phys. Rev. Lett. \textbf{67}, 181 (1991).
\bibitem{palm}
J. J. McClelland, R. E. Scholten, E. C. Palm, and R. J. Celotta, Science \textbf{262}, 877 (1993).
\bibitem{4}
F. Riehle, \textit{Frequency Standards. Basics and Applications} (Wiley, Weinheim, 2004).
\bibitem{1a}
R. Islam, E. Edwards, K. Kim, S. Korenblit, C. Noh, H. Carmichael, G.-D. Lin, L.-M. Duan, C.-C. Joseph Wang, J. Freericks, and C. Monroe, Nat. Commun. \textbf{2}, 377 (2011).
\bibitem{10} 
M. H. G. de Miranda, A. Chotia, B. Neyenhuis, D. Wang, G. Qu\'{e}m\'{e}ner, S. Ospelkaus, J. L. Bohn, J. Ye, and D. S. Jin, Nat. Phys. \textbf{7}, 502 (2011).
\bibitem{14a}
S. Ejtemaee and P. C. Haljan, Phys. Rev. A \textbf{87}, 051401 (2013).
\bibitem{15a}
K. Pyka, J. Keller, H. L. Partner, R. Nigmatullin, T. Burgermeister, D. M. Meier, K. Kuhlmann, A. Retzker, M. B. Plenio, W. H. Zurek, A. del Campo, and T. E. Mehlst\"{a}ubler, Nat. Commun. \textbf{4}, 2291 (2013).
\bibitem{17a} 
M. B. Plenio and A. Retzker, Annalen der Physik \textbf{525}, A159 (2013).
\bibitem{18a}
A. Bylinskii, D. Gangloff, and V. Vuletic, Science \textbf{348}, 1115 (2015). 
\bibitem{Schawlow}
T. W. H\"{a}nsch and A. L. Schawlow, opt. Commun \textbf{13}, 68 (1975).
\bibitem{note}
Laser cooling of trapped ions was proposed at about the same time; D. J. Wineland and H. G. Dehmelt, Bull. Am. Phys. Soc. \textbf{20}, 637 (1975).
\bibitem{adams}
C. S. Adams and E. Riis, Prog. Quant Electron \textbf{21}, 1 (1997).
\bibitem{lett}
P. D. Lett, R. N. Watts, C. I. Westbrook, W. D. Phillips, P. L. Gould, and H. J. Metcalf, Phys. Rev. Lett. \textbf{61}, 169 (1988).
\bibitem{Dalibard}
J. Dalibard and C. Cohen-Tannoudji, J. Opt. Soc. Am. B \textbf{6}, 2023 (1989).
\bibitem{Ungar}
P. J. Ungar, D. S. Weiss, E. Riis, and S. Chu, J. Opt. Soc. Am. B \textbf{6}, 2058 (1989).
\bibitem{weiss2}
D. S. Weiss, E. Riis, Y. Shevy, P. Jeffrey Ungar, and S. Chu, J. Opt. Soc. Am. B \textbf{6}, 2072 (1989).
\bibitem{Hoogerland}
M. D. Hoogerland, H. C. W. Beijerinck, K. A. H. van Leeuwen, P. van der Straten, and H. J.  Metcalf, Europhys Lett. \textbf{19}, 669 (1992).
\bibitem{Kasevich2}
M. Kasevich and S. Chu, Phys. Rev. Lett. \textbf{69}, 1741 (1992).
\bibitem{Bergeman}
T. Bergeman, Phys. Rev. A \textbf{48}, 3425 (1993).
\bibitem{Metcalf}
H. J. Metcalf and P. Van der Straten, \textit{Laser cooling and trapping} (Springer Science and Business Media, New York, 2012).
\bibitem{Arimondo}
E. Arimondo, \textit{Progress in Optics XXXV} (North-Holland, Amsterdam, 1996).
\bibitem{eit}
S. Harris, Phys. Today. \textbf{50}, 36 (1997).
\bibitem{scully}
M. O. Scully and M. S. Zubairy, \textit{Quantum optics} (Cambridge University Press, Cambridge, 1997).
\bibitem{hariss2}
S. E. Harris and A. V. Sokolov, Phys. Rev. Lett. \textbf{81}, 2894 (1998).
\bibitem{harris}
L. V. Hau, S. E. Harris, Z. Dutton, and C. H. Behroozi, Nature \textbf{397}, 594 (1999).
\bibitem{kash}
M. M. Kash, V. A. Sautenkov, A. S. Zibrov, L. Hollberg, G. R. Welch, M. D. Lukin, Y. Rostovtsev, E. S. Fry, and M. O. Scully, Phys. Rev. Lett. \textbf{82}, 5229 (1999).
\bibitem{budker} 
D. Budker, D. F. Kimball, S. M. Rochester, and V. V. Yashchuk, Phys. Rev. Lett. \textbf{83}, 1767 (1999).
\bibitem{Aspect}
A. Aspect, E. Arimondo, R. E. A. Kaiser, N. Vansteenkiste, and C. Cohen-Tannoudji, Phys. Rev. Lett. \textbf{61}, 826 (1988).
\bibitem{15c}
K.-J. Boller, A. Imamo\u{g}lu, and S. E. Harris, Phys. Rev. Lett. \textbf{66}, 2593 (1991).
\bibitem{16c}
G. Morigi, J. Eschner, and C. H. Keitel, Phys. Rev. Lett. \textbf{85}, 4458 (2000).
\bibitem{he}
P. He, P. M. Tengdin, D. Z. Anderson, A. M. Rey, and M. Holland, Phys. Rev. A. \textbf{95}, 053403 (2017).
\bibitem{lukin}
M. D. Lukin, S. F. Yelin, M. Fleischhauer, and M. O. Scully, Phys. Rev. A \textbf{60}, 3225 (1999).
\bibitem{chen2001}
Y. C. Chen, Y. A. Liao, H. Y. Chiu, J. J.  Su, and  A. Y. Ite, Phys. Rev. A \textbf{64}, 053806 (2001).
\bibitem{ye}
C. Y. Ye, A. S. Zibrov, Y. V. Rostovtsev, and M. O. Scully, Phys. Rev. A \textbf{65}, 043805 (2002).
\bibitem{yelin}
S. F. Yelin, V. A. Sautenkov, M. M. Kash, G. R. Welch, and M. D. Lukin, Phys. Rev. A \textbf{68}, 063801 (2003).
\bibitem{mahmoudi}
M. Mahmoudi, R. Fleischhaker, M. Sahrai M, and J. Evers, J. Phys. B: At. Mol. Opt. Phys. \textbf{41}, 025504 (2008).
\bibitem{zoller}
C. Gardiner and P. Zoller, \textit{Quantum Noise} (Springer Science and Business Media, New York, 2004).
\bibitem{mckay}
D. C. McKay and B. DeMarco, Rep. Prog. Phys. \textbf{74}, 054401 (2011).
\bibitem{simon}
M. Xu, S. B. J\"{a}ger, S. Schütz, J. Cooper, G. Morigi, and  M. J. Holland, Phys. Rev. Lett. \textbf{116}, 153002 (2016).
\bibitem{cohen}
C. Cohen-Tannoudji and D. Gu\'{e}ry-Odelin, \textit{Advances in atomic physics: an overview} (World Scientific, Singapore, 2011).
\bibitem{molmer}
K. Mølmer, Y. Castin, and J. Dalibard, J. Opt. Soc. Am. B \textbf{10}, 524 (1993).
\bibitem{castin}
J. Dalibard, Y. Castin, K. and Mølmer, Phys. Rev. Lett. \textbf{68}, 580 (1992).
\bibitem{childress}
M. D. Lukin and L. Childress. Modern Atomic and Optical Physics II Lecture. Notes. available online at http://ultracold.jqi.umd.edu/atomic-physics-721/, 2005.
\bibitem{chang}
D. E. Chang, V. Vuleti\'{c}, and M. D. Lukin, Nat. Photonics \textbf{8}, 685 (2014).
\bibitem{gross}
M. Gross and S. Haroche, Phys. Rep. \textbf{93}, 301 (1982).
\bibitem{castin2}
Y. Castin, J. I. Cirac, and M. Lewenstein, Phys. Rev. Lett. \textbf{80}, 5305 (1998).
\end{thebibliography}
\end{document}